# Thermal transport in the anisotropic Heisenberg chain with S = 1/2 and nearest-neighbor interactions

D. L. Huber

Department of Physics, University of Wisconsin-Madison,

Madison, WI 53706

## Abstract

The purpose of this note is to connect early work on thermal transport in spin-1/2 Heisenberg chains with uniaxial exchange anisotropy and nearest-neighbor interactions that was based on a moment analysis of the Fourier transform of the energy density correlation function with subsequent studies that make use of thermal current correlation functions.

Corresponding author:

D. L. Huber, Physics Department, University of Wisconsin-Madison, 1150 University Avenue, Madison, WI 53706, USA

E-mail huber@src.wisc.edu; phone: 608-265-4035; fax 608-265-2334

## I. Introduction

Recent experimental [1] and theoretical studies [2,3] have focused renewed attention on the unusual thermal transport properties of anisotropic spin-1/2 Heisenberg chains with nearest-neighbor interactions. The first theoretical investigations of the transport involved an analysis of the second and fourth moments of the energy density correlation function [4,5]. In Ref. 4, it was found that the thermal transport was non-diffusive in the infinite temperature limit for all values of the anisotropy (excluding the Ising limit). The calculations in Ref. 5 extended the results of Ref. 4 for the isotropic chain to finite temperatures. Subsequently, Th. Niemeijer and H. A. W. van Vianen [6] showed that the thermal current operator for the anisotropic, spin-1/2 Heisenberg chain with nearest-neighbor interactions was a constant of the motion in the absence of a field. They also found that in the presence of a field, the thermal current operator was the sum of two parts; one of which was identical to the zero-field thermal current, and thus a constant of the motion, while the other, proportional to the field, was a constant of the motion only for the XY (non-interacting fermion) model. The purpose of this note is to show the connection between the analysis of the diffusion constant in Refs. 4 and 5, and the thermal current results presented in Ref. 6.

## II. Analysis

The connection between the energy density correlation function and the thermal current density correlation function is through the continuity equation [7]. Denoting the Hamiltonian density of the spin system by $h(z,t)$ and the thermal current density operator by $j(z,t)$, the continuity equation is expressed by the familiar form

$$\partial h / \partial t + \partial j / \partial z = 0 \tag{1}$$

Writing the spatial transforms as $\tilde{h}(q,t)$ and $\tilde{j}(q,t)$, the equivalent continuity equation takes the form

$$\partial \tilde{h}(q,t) / \partial t + i q \tilde{j}(q,t) = 0 \tag{2}$$

Equation (2) allows one to establish a connection between the Fourier transform of the energy density correlation function and the transform of the corresponding thermal current density correlation function. In the infinite temperature limit, the equation takes the form

$$\omega^2 E(q,\omega) = q^2 J(q,\omega) \tag{3}$$

where

$$E(q,\omega) = \frac{1}{2\pi} \int_{-\infty}^{\infty} dt <\tilde{h}(q,t)\tilde{h}(-q,0)> \exp[i\omega t] \tag{4}$$

and

$$J(q,\omega) = \frac{1}{2\pi} \int_{-\infty}^{\infty} dt <\tilde{j}(q,t)\tilde{j}(-q,0)> \exp[i\omega t] \tag{5}$$

As noted, the analysis of Refs. 4 and 5 is based on the moments of the energy density function. In normalized form, these moments are given by

$$<\omega^{2n}(q)> = \int_{-\infty}^{\infty} \omega^{2n} E(q,\omega) d\omega / \int_{-\infty}^{\infty} d\omega E(q,\omega) \tag{6}$$

Thus Eq.(3) leads to

$$<\omega^2(q)> = q^2 \int_{-\infty}^{\infty} d\omega J(q,\omega) / \int_{-\infty}^{\infty} d\omega E(q,\omega) \qquad (7)$$

and

$$<\omega^4(q)> = q^2 \int_{-\infty}^{\infty} \omega^2 J(q,\omega) d\omega / \int_{-\infty}^{\infty} d\omega E(q,\omega) \qquad (8)$$

The approach followed in Ref. 6 was based on the (total) thermal current operator, which is equivalent to $\tilde{j}(0,t)$, and its correlation function $J(0,\omega)$. We return to this point in the following section.

## III. Discussion

In Ref. 4, the thermal transport was analyzed in terms of the moment fluctuation ratio as proposed by Bennett [8]. The thermal moment fluctuation ratio has the form

$$R_E(q) = <\omega^2(q)>^2 / [<\omega^4(q)> - <\omega^2(q)>^2] \qquad (9)$$

When $R_E(q) << 1$, the transport is diffusive and when $R_E(q) >> 1$ weakly damped wave propagation dominates. In the case of thermal transport in the anisotropic, spin-1/2 Heisenberg chain, it was found that the fourth moment varied as $q^4$ in the small-q limit, so that $R_E(q)$ approaches a finite value as $q \to 0$, thus ruling out thermal diffusion. The small-q limit of the fourth moment has the form

$$<\omega^4(q)> = q^2 \int_{-\infty}^{\infty} \omega^2 J(0,\omega) d\omega / \int_{-\infty}^{\infty} d\omega E(0,\omega) + O(q^4) \qquad (10)$$

The vanishing of the quadratic term is a direct consequence of the of the thermal current operator being a constant of the motion so that $J(0,\omega)$ is proportional to $\delta(\omega)$.

As noted in the Introduction, the authors of Ref. 6 found that in the presence of an applied field, the thermal current operator was the sum of two terms: a term identical to the current operator in the absence of the field, and thus a constant of the motion, and a second term, proportional to the field, that was a constant of the motion only in the limit where the Hamiltonian corresponds to the XY model with the field directed along the z-axis. Except in the XY limit, the Fourier transform of the correlation function of the thermal current operator is predicted to have the 'Drude' form [2]

$$J(0,\omega) = A\delta(\omega) + B(\omega) \qquad (11)$$

where $B(\omega)$ is non-singular.

Finally, we note that infinite temperature analysis presented here can be extended to finite temperatures by replacing the correlation functions by relaxation functions [9, 10] as was done in Ref. 5. The analysis of the moments of the Fourier transform of the relaxation function is similar to what was done above with the correlation functions, and the conclusions are the same.

## Acknowledgments

The author would like to thank W. Brenig and X. Zotos for reprints of their papers.